\documentclass[aps,pra,twocolumn,showpacs,preprintnumbers,amsmath,amssymb]{revtex4-2}
\usepackage{soul}
\usepackage{amsmath}
\usepackage{esvect}

\usepackage{natbib}
\usepackage{graphicx}
\usepackage{color}
\usepackage{tabularx}
\usepackage{multirow}

\usepackage{wasysym}

\usepackage{amssymb}

\usepackage[utf8]{inputenc}

\usepackage{float}

\usepackage{subfigure}

\usepackage{float}

\usepackage{hyperref}

\usepackage{cancel}

\usepackage{xcolor}

\usepackage{setspace}
\usepackage{soul}
\usepackage{wrapfig}

\definecolor{burntorange}{rgb}{0.8, 0.33, 0.0}

\begin{document}

\title{A nonlinear frequency shift caused by asymmetry\\
of the coherent population trapping resonance: a generalization}

\author{E.\,A.~Tsygankov$^{1}$}
\email[]{tsygankov.e.a@yandex.ru}
\author{D.\,S.~Chuchelov$^{1}$}
\author{M.\,I.~Vaskovskaya$^{1}$}
\author{V.\,V.~Vassiliev$^{1}$}
\author{S.\,A.~Zibrov$^{1}$}
\author{V.\,L.~Velichansky$^{1}$}
\affiliation{1. P.\,N. Lebedev Physical Institute of the Russian Academy of Sciences,\\
Leninsky Prospect 53, Moscow, 119991 Russia}

\newcommand{\red}{\textcolor{red}}
\newcommand{\blue}{\textcolor{blue}}

\begin{abstract}
We investigate the coherent population trapping resonance induced by a polychromatic optical field with an asymmetric spectrum, i.e., whose sidebands equidistant from the carrier have unequal powers. A situation is considered where a modulation is used to provide the \hbox{so-called} in-phase and quadrature signals in the optical field's transmission, which are used for stabilization of the local oscillator's frequency in chip-scale atomic clocks. In a general case, the frequencies of the signals nonlinearly depend on the optical field's intensity due to the asymmetry of the resonance. In this work, we demonstrate a) that this effect stems from a multi-peak structure of the resonance; b) a linear dependence of the frequency on the optical field's intensity at high modulation values; c) that the regime of the resolved structure has more advantages than suppression of the frequency shift due to the spectrum asymmetry.
\end{abstract}

\maketitle

\section{Introduction}

The most basic elements of chip-scale atomic clocks are a miniature glass cell filled with alkali-metal atoms and a buffer gas, and a vertical-cavity surface-emitting laser~\cite{doi:10.1063/1.5026238}. A microwave modulation of the latter's injection current is used to obtain a proper optical field and induce the coherent population trapping (CPT) resonance. The modulation frequency $\Omega$ is usually corresponds to half of alkali-metal atoms' ground-state hyperfine splitting $\omega_g$. In this case the first sidebands (often called as ``resonant components'') of the laser field are tuned to the absorption line. The transmission depends on the difference $2\Omega-\omega_g=\delta$ and reaches a maximal level when $\delta=0$. The resonance itself cannot be used for stabilization of the local oscillator frequency since it is an even function of two-photon detuning $\delta$. Therefore, a modulation of frequency $\Omega$ or its phase can be implemented, $\Omega\rightarrow\Omega+m\omega_m\cos{\omega_mt}$, which provides oscillations of the light absorption at multiples of $\omega_m$. Here $m$ is the modulation index. Amplitudes of the odd harmonics are proportional to $\delta$ having a dispersive shape~\cite{1402-4896-93-11-114002} and therefore they are suitable for the stabilization loop. Usually, oscillations at frequency $\omega_m$ are used, their parts $\propto\cos{\omega_mt}$ and $\propto\sin{\omega_mt}$ are called as ``in-phase'' and ``quadrature'' signals, respectively. Their mixture providing an optimal slope steepness is often taken as the error signal for stabilization of the local oscillator frequency~\cite{ben2007optimization,kahanov2008dependence}.

In most cases, the first sidebands have unequal powers due to a nonlinear interaction of spectral components in the active medium of the diode laser~\cite{Tsygankov:22}. In a general situation, this leads to the asymmetry of the CPT resonance: it becomes neither an even nor an odd function of the two-photon detuning. Authors of work~\cite{yudin2023frequency} have recently demonstrated that the asymmetry leads to a nonlinear dependence of the in-phase and quadrature signals frequencies on the optical field intensity. However, we have demonstrated this several years ago in our work~\cite{https://doi.org/10.48550/arxiv.2012.13731}, in which the double $\Lambda$-scheme of levels and a polychromatic optical field with an asymmetric spectrum were considered. Then we have obtained analytical expressions for the shift $\delta_{as}$ of the signals frequencies occurring due to the resonance asymmetry, which are not proportional to the laser field intensity $I$ in contrast to the light shift. This feature hinders effectiveness of techniques for suppression of the microwave transition frequency's light shift (further we will use the phrase ``the light shift'' for brevity) based on modulation of the total optical field intensity in the following way. When condition $\partial(\delta_{as}+\delta_0)/\partial I=0$ is fulfilled, where $\delta_0$ is the light shift, the total shift itself is not equal to zero, $\delta_{as}+\delta_0\neq0$. A situation occurs when there is no response in frequencies of signals to variations of the optical field intensity, but they remain displaced from the microwave transition frequency. The values of their shifts are obviously depend on parameters of the laser radiation and the medium since they determine a degree of the asymmetry. Therefore, despite one ensures the situation $\partial(\delta_{as}+\delta_0)/\partial I=0$, frequencies of the signals can undergo random walks which reduce clocks' performance.

We remind here that the light shift is considered as the main factor limiting the long-term frequency stability of chip-scale atomic clocks~\cite{Knappe:01,Vanier2005,Gerginov:06,SHAH201021,2015NCimR..38..133G}; the standard approach for its suppression consist in finding a balance in the spectrum: components with index $|k|\geq2$ and $|k|\leq1$ induce the light shift of the opposite sign~\cite{827437,Zhu,doi:10.1063/1.2360921} (but such a situation is possible if the pressure of a buffer gas is not too high~\cite{BGpaper}). Nonlinear dependence of the signals frequencies on the laser field intensity makes a problem of controlling the spectrum just more complicated.

In this work, we demonstrate that nonlinear dependence of $\delta_{as}$ on the optical field intensity is a consequence of an unresolved multi-peak structure of the CPT resonance and show a linear dependence of $\delta_{as}$ on $I$ at high values of $\omega_m$ as proposed in~\cite{yudin2023frequency}.

\section{Theory}
\label{Theory}

To describe the coherent population trapping resonance, at least a $\Lambda$-system of levels is required. The excited-state level $|e\rangle$ should be coupled with ground ones $|a\rangle$ and $|b\rangle$, which we considered here as non-degenerate, $\omega_a-\omega_b=\omega_g$. In our case the coupling is due to electric-dipole transitions induced by bichromatic optical field
{\small
\begin{equation}
\mathcal{E}(t)=-\dfrac{1}{2}\left\{\mathcal{E}_{-1}e^{-i\left[(\omega_0-\Omega)t-\varphi(t)\right]}-\mathcal{E}_1e^{-i\left[(\omega_0+\Omega)t+\varphi(t)\right]}\right\}+c.c.,
\end{equation}}

\noindent where $\varphi(t)=m\sin{\omega_m}t$ is the modulation needed for producing in-phase and quadrature signals. The frequency $\omega_0$ is taken to be equal to half-sum of the involved transitions $\omega_e-(\omega_a+\omega_b)/2$, where $\omega_e\gg\omega_{a,b}$. I.e., the short- and long-wavelength spectral component are resonant to transition $|a\rangle\rightarrow|e\rangle$ and $|b\rangle\rightarrow|e\rangle$, correspondingly. The dipole moments of both transitions are taken equal and real. The frequency $\Omega$ is close to $\omega_g/2$, and the difference gives two-photon detuning $\delta$.

It is known, that we can arrive at the following set of equations for the density matrix elements using the rotating wave approximation, assuming the low saturation regime and adiabatically eliminating the excited state:
\begin{subequations}
\begin{equation}
\rho_{ee}=\dfrac{2}{\gamma\Gamma}\left\{V^2_{-1}\rho_{aa}+V^2_1\rho_{bb}-2V_{-1}V_1\text{Re}\left[\tilde{\rho}_{ab}(t)\right]\right\},
\end{equation}
\begin{equation}
\left[i\frac{\partial}{\partial t}+2\left(\tilde{\delta}+m\omega_m\cos{\omega_m t}\right)+i\tilde{\Gamma}_g\right]\tilde{\rho}_{ab}=i\dfrac{V_{-1}V_1}{\Gamma}.
\label{rhoab}
\end{equation}
\end{subequations}

In the equations above, we imply for the Rabi frequencies that $V_{-1}=d\mathcal{E}_{-1}/2\hbar=V_1=d\mathcal{E}_1/2\hbar=V$ and $\rho_{aa}=\rho_{bb}\simeq1/2$, but retain the lower indices for convenience, $\gamma$ and $\Gamma$ is the decay rate of the excited-state population and of the optical coherences, respectively. The ground-state relaxation rate $\tilde{\Gamma}_g=\Gamma_g+(V^2_{-1}+V^2_1)/\Gamma$ accounts for the resonance power broadening. The light shift $\delta_0$ can be accounted in detuning $\tilde{\delta}=\delta-\delta_0$, but it is not the object of our interest here. Finally, we specially note that $\tilde{\rho}_{ab}$ is not the coherence itself, but its slowly varying amplitude: $\rho_{ab}=\tilde{\rho}_{ab}e^{-2i\Omega t}$.

Equation~\eqref{rhoab} can be straightforwardly integrated or be solved numerically. However, these approaches do not give understanding about influence of the modulation on the CPT resonance structure. Instead, we will use the replacement $\tilde{\rho}_{ab}\rightarrow\bar{\rho}_{ab}e^{2im\sin{\omega}_mt}$ to make like a step behind in derivation of the equations. This gives
{\small
\begin{subequations}
\begin{equation}
\rho_{ee}=\dfrac{2}{\gamma\Gamma}\left\{V^2_{-1}\rho_{aa}+V^2_1\rho_{bb}-2V_{-1}V_1\text{Re}\left[e^{2im\sin{\omega}_mt}\bar{\rho}_{ab}(t)\right]\right\},
\end{equation}
\begin{equation}
\left(i\frac{\partial}{\partial t}+2\tilde{\delta}+i\tilde{\Gamma}_g\right)\bar{\rho}_{ab}=i\dfrac{V_{-1}V_1}{\Gamma}e^{-2im\sin{\omega}_mt}.
\label{rhoab2}
\end{equation}
\end{subequations}}

The equations above now can be treated via the Fourier series expansion of $\bar{\rho}_{ab}$ over the frequency $\omega_m$ and Jacobi–Anger one for $e^{\mp2im\sin{\omega_m}t}$. At first, for the zeroth harmonic of $\rho_{ee}$, we have
{\small
\begin{equation}
\dfrac{1}{T}\int^{T}_{0}\rho_{ee}dt=\dfrac{V^2}{\gamma\Gamma}\left[1-2\tilde{\Gamma}_g\dfrac{V^2}{\Gamma}\sum^{\infty}_{k=-\infty}
\dfrac{J^2_k(2m)}{(2\tilde{\delta}+k\omega_m)^2+\tilde{\Gamma}^2_g}\right],
\label{zeroth}
\end{equation}}

\noindent where $T=2\pi/\omega_m$ and we used relations $V_{-1}=V_1$ and $\rho_{aa}=\rho_{bb}=1/2$, $J_k(\cdot)$ is the Bessel function of the first kind of the index $k$.

\begin{figure}[b]
\centering
\includegraphics[width=\columnwidth]{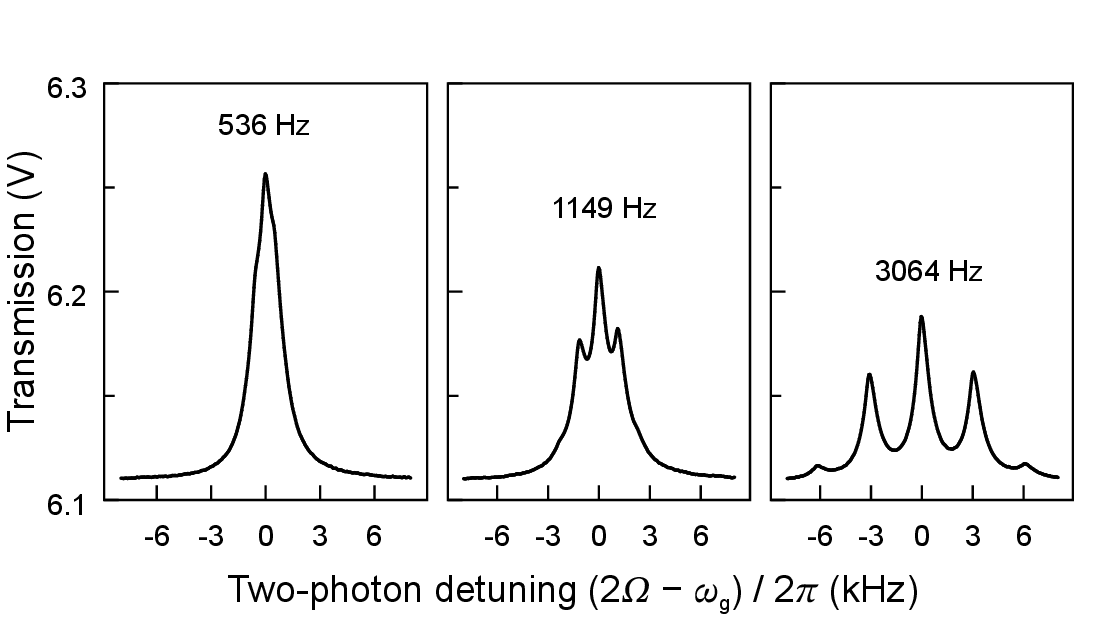}
\caption{The experimental multi-peak structure of the CPT resonance (zeroth harmonic of the optical field transmission by an atomic cell over frequency $\omega_m$) for $\omega_m/2\pi$ equal to $536$, $1149$, and $3064$~Hz; see Section~\ref{Exp} for details. The value of modulation index $m$ was fixed and equal to $0.65$.}
\label{MultiPeak}
\end{figure}

Equation~\eqref{zeroth} demonstrates that the modulation provides multi-peak structure of the CPT resonance, which becomes resolved with growth of $\omega_m$; see Fig.~\ref{MultiPeak}. This feature can be understand as follows. In contrast to the standard situation, the optical field prepares the ground-state coherence not only at the frequency $2\Omega$ (difference of resonant spectral components frequencies), but at frequencies $2\Omega+k\omega_m$. 
When \hbox{$2\Omega+k\omega_m=\omega_g$}, amplitude of corresponding oscillations of the ground-state coherence reaches maxima providing minimum in the absorption for the probing process.

\begin{figure}[t]
\centering 
\includegraphics[width=\columnwidth]{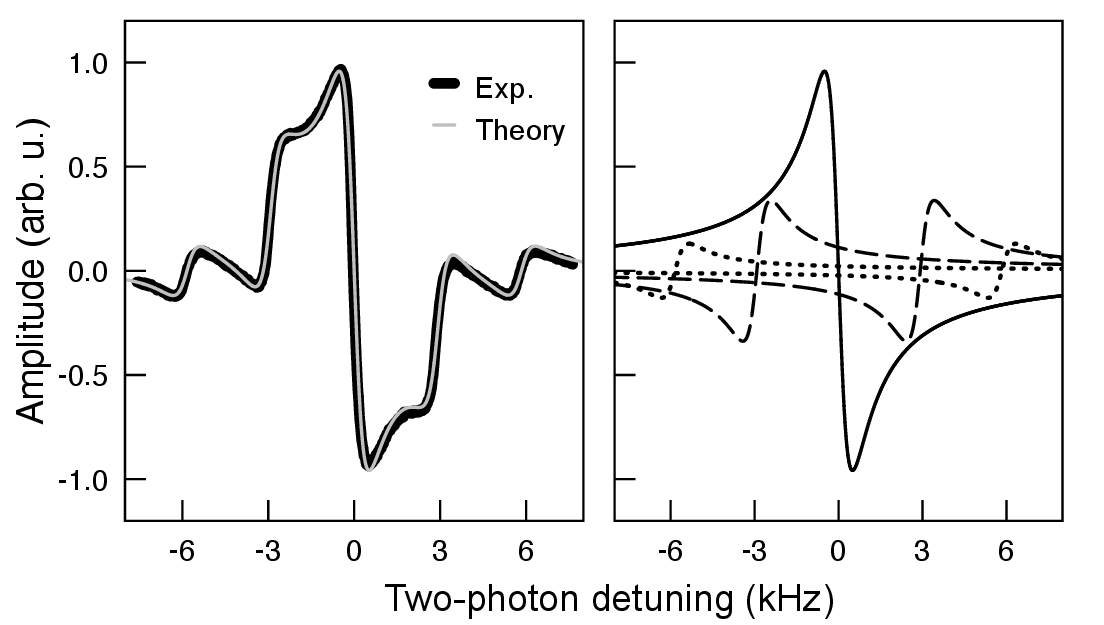}
\caption{Left: the experimental error signal (first harmonic of the optical field transmission by an atomic cell over frequency $\omega_m$) for \hbox{$\omega_m/2\pi=3064$~Hz}, $m=0.65$ (see Section~\ref{Exp} for details), and the fit made via Eq.~\eqref{quadr}. The value of $\tilde{\Gamma}_g/2\pi$ in the fit is $\simeq520$~Hz. Right: individual curves of the error signal from the left part of the figure plotted according to terms in the sum of the same formula. The central one is for $k=0$, the large and fine dotted curves are for $k=\pm1$, $k=\pm2$ correspondingly.}
\label{Error-signal}
\end{figure}

Considering the in-phase and quadrature signals, their shape can be relatively complicated at moderate values of $m$ and $\omega_m$. This feature also stems from the fact that they have a multi-component structure. For example, amplitude of the quadrature signal $A_Q$ is determined by the function
\begin{equation}
A_Q\propto\sum^{\infty}_{k=-\infty}(2\bar{\delta}+k\bar{\omega}_m)\dfrac{J_k(2m)\left[J_{k-1}(2m)-J_{k+1}(2m)\right]}{\left(2\bar{\delta}+k\bar{\omega}_m\right)^2+1},
\label{quadr}
\end{equation}

\noindent where $\bar{\delta}=\tilde{\delta}/\tilde{\Gamma}_g$, $\bar{\omega}_m=\omega_m/\tilde{\Gamma}_g$ (parameters are normalized on the ground-state coherence relaxation rate), which demonstrates that there are dispersive-shape curves that have zero-crossing point at frequencies corresponding to peaks in the mean in time absorption; see Fig.~\ref{Error-signal}.

There are oscillations of $\rho_{ee}$ not only at frequency $\omega_m$, but also at it multiples, $k\omega_m$. They can be understand as a result of probing the coherence oscillations at frequency \hbox{$2\Omega+s\omega_m$} by the product of optical field's components oscillating at frequencies \hbox{$2\Omega+s\omega_m\mp k\omega_m$}. Therefore, for example, there are terms \hbox{$\propto J_{k}(2m)J_{k-1}(2m)$}, \hbox{$\propto J_{k}(2m)J_{k+1}(2m)$} in Eq.~\eqref{quadr}. Function, determining oscillations of $\rho_{ee}$ at $2\omega_m$, will contain products \hbox{$J_{k}(2m)J_{k-2}(2m)$} and \hbox{$J_{k}(2m)J_{k+2}(2m)$}, etc.

Finally, when $V_{-1}\neq V_1$, and frequencies of spectral components are detuned from transitions $|a\rangle\rightarrow|e\rangle$ and $|b\rangle\rightarrow|e\rangle$, the CPT resonance becomes asymmetric. In this case, the side-peaks of $\dfrac{1}{T}\int^{T}_{0}\rho_{ee}dt$ pull the frequency of the central one, and the similar situation occurs for the in-phase and quadrature signals. As far as under the asymmetry $\rho_{aa}\neq\rho_{bb}$, equations for ground-state populations are required to describe the CPT resonance. Fourier amplitudes of the coherence and populations are coupled and there is no general analytical solution of equations. However, for $a\ll1$, the expressions describing frequency shifts of the signals can be obtained. Specifically, for the zero-crossing point of the signals sum, we have the following shift:
\begin{equation}
\delta_{as}\propto\dfrac{V^2_{-1}-V^2_1}{\Gamma}\dfrac{(1+\bar{\omega}_m)(1+\bar{\omega}^2_m)}{1+\bar{\omega}_m(3+\bar{\omega}^2_m)}.
\label{pulling}
\end{equation}

This function falls with growth of $\bar{\omega}_m$ from the zero, reaches a minimum at $\bar{\omega}_m\simeq0.47$ and after that grows. For \hbox{$\bar{\omega}_m\gg1$} the shift has the same value as for \hbox{$\bar{\omega}_m\ll1$}, i.e., this function demonstrates a behavior typical for the frequency pulling. Also, $\delta_{as}$ linearly depends on the optical field intensity in these cases since it is proportional to \hbox{$(V^2_{-1}-V^2_1)/\Gamma$}. In contrast, $\delta_{as}$ is a nonlinear function of the optical field intensity at moderate values of $\bar{\omega}_m$ as far as $\tilde{\Gamma}_g$ contains the power broadening.

\section{Experiment}
\label{Exp}

\begin{figure}[b]
\centering 
\includegraphics[width=\columnwidth]{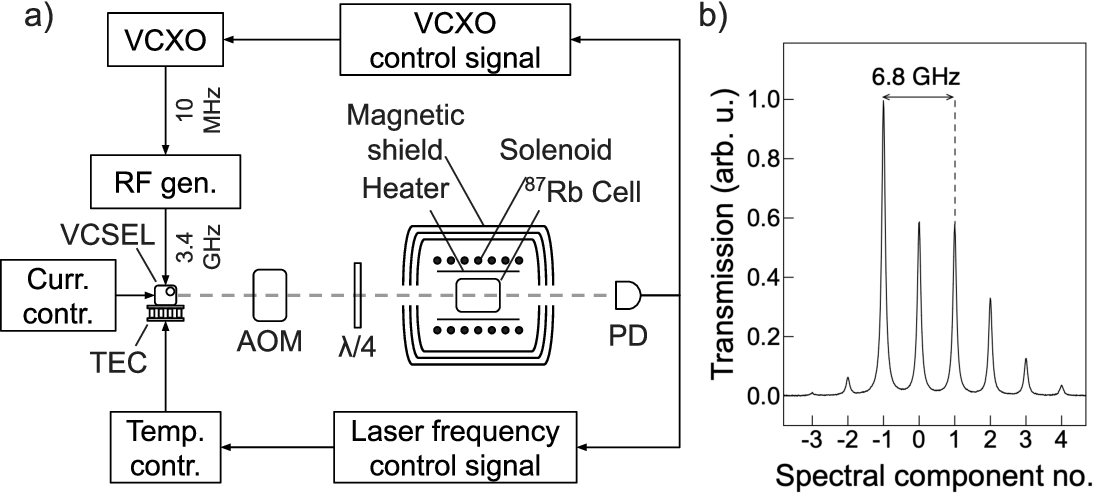}
\caption{a) Scheme of the experimental setup. b) Spectrum of the laser radiation used in the experiment.}
\label{ExpSetup}
\end{figure}

The experimental setup is schematically shown in Fig.~\ref{ExpSetup}a. We used a single-mode vertical-cavity surface-emitting laser VCSEL generating at \hbox{$\simeq795$~nm}. The DC and RF components of the injection current were fed to the laser via a bias tee. The modulation frequency $\Omega/2\pi$ was close to $3.417$~GHz, and the first sidebands of the polychromatic optical field were tuned to transitions $F_g=2\rightarrow F_e=2$, $F_g=1\rightarrow F_e=2$ of the $^{87}$Rb D$_1$~line. The power of the RF field and the injection current value were set to provide a significant difference between powers of the resonant spectral components (see inset in Fig.~\ref{ExpSetup}b) and cause a noticeable asymmetry of the CPT resonance. A quarter-wave plate was used to form the CPT resonance in \hbox{$\sigma^+$-$\sigma^+$} scheme. The diameter of the laser beam was $3$~mm. The laser wavelength was stabilized by a feedback loop that controls the temperature of the laser diode.

A cylindrical atomic cell ($8$~mm diameter, $15$~mm length, $0.7$~mm wall thickness) filled with isotopically enriched $^{87}$Rb and Ne at pressure of $90$~Torr was under study. The atomic cell was placed in a longitudinal magnetic field of $0.02$~G to separate the metrological microwave transition from magneto-sensitive ones at sublevels $m_{F_g}=\pm1$. The temperature of the atomic cell was maintained close to $65$~$^{\circ}$C with an accuracy of $0.01$~$^{\circ}$C. The cell, the heater, and the solenoid were placed in a three-layer $\mu$-metal magnetic shield, providing an over $500$-fold suppression of the laboratory magnetic field.

To generate the error signal for stabilization of the VCXO frequency, the RF signal frequency was modulated: \hbox{$j(t)=j_{\text{DC}}+j_{\text{RF}}\cos{(\Omega t+m\sin{\omega_mt})}$}. The in-phase and quadrature signals were registered using synchronous detection of the transmission signal of the atomic cell at the modulation frequency $\omega_m/2\pi$.

Fig.~\ref{MultiPeak} shows the experimentally registered CPT resonance. The RF frequency was scanned at a rate of $1$~kHz/s around $3.417359$~GHz. The obtained signal of the cell transmission was low-pass filtered ($10$~Hz) and averaged over $5$~scans. The laser radiation power was $60$~$\mu$W. Fig.~\ref{MultiPeak} demonstrates evolution of the laser field transmission with growth of the modulation frequency $\omega_m$ while maintaining the modulation index $m=0.65$. When value of $\omega_m/2\pi$ exceeds the ground-state relaxation rate, the multi-peak structure becomes resolved. In the experiment we clearly observed five separated peaks for modulation frequency $\omega_m/2\pi=3064$~Hz. The amplitudes $A_k$ of the registered peaks are determined by the squares of the Bessel functions of the corresponding index: for the central peak $A_0\propto J^2_0(2m)$, for the first side peaks $A_{\pm1}\propto J^2_1(2m)$, etc., as follows from Eq.~\eqref{zeroth}. It can be seen that all peaks have the same type of asymmetry: the left slope is steeper than the right. Therefore, the frequency of the central peak is pulled.
\begin{figure}[b]
\centering 
\includegraphics[width=\columnwidth]{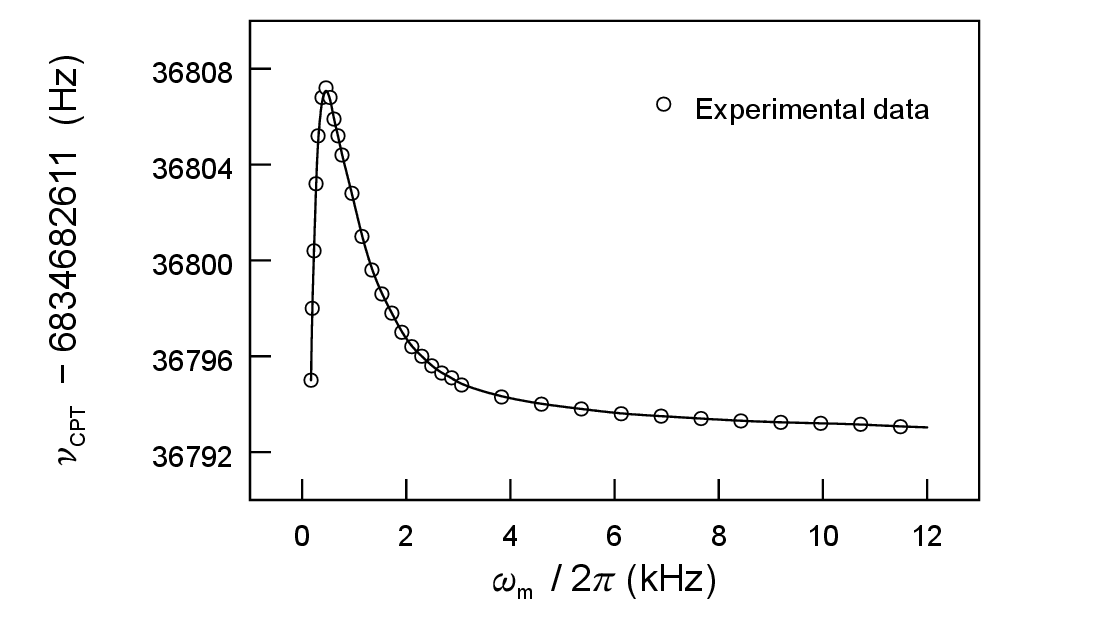}
\caption{Dependence of the error-signal frequency on $\omega_m/2\pi$. Circles demonstrate the experimental data, the solid line is the guide for eyes. The value of modulation index $m$ was fixed and equal to $0.65$.}
\label{Pulling}
\end{figure}

Fig.~\ref{Pulling} shows the dependence of the CPT resonance frequency on $\omega_m$ for a fixed value of $m=0.65$. The VCXO frequency was stabilized at a value corresponding to the zero-crossing point of the error signal and was measured with a frequency counter. When the modulation frequency is smaller than individual width of the peaks, the in-phase signal has a steeper slope than the quadrature signal. In the opposite case the steepness of the in-phase signal drops and the quadrature signal dominates. For each value of $\omega_m$ the phase of synchronous detection was set to maximize the slope, therefore a mixture of in-phase and quadrature signals was used as the error signal. The initial increase of the frequency at small values of $\omega_m$ is associated with pulling of the central peak towards the high-frequency side peaks. The maximum shift of the error-signal frequency, caused by the pulling effect, reaches a value of $14$~Hz. It is achieved at $\omega_m/2\pi\simeq450$~Hz, which is close to the half-width of the CPT resonance for $m=0$ and is in good agreement with the theoretical prediction. At modulation frequencies significantly higher than widths of the peaks the effect of frequency pulling is small.
\begin{figure}[b]
\centering 
\includegraphics[width=\columnwidth]{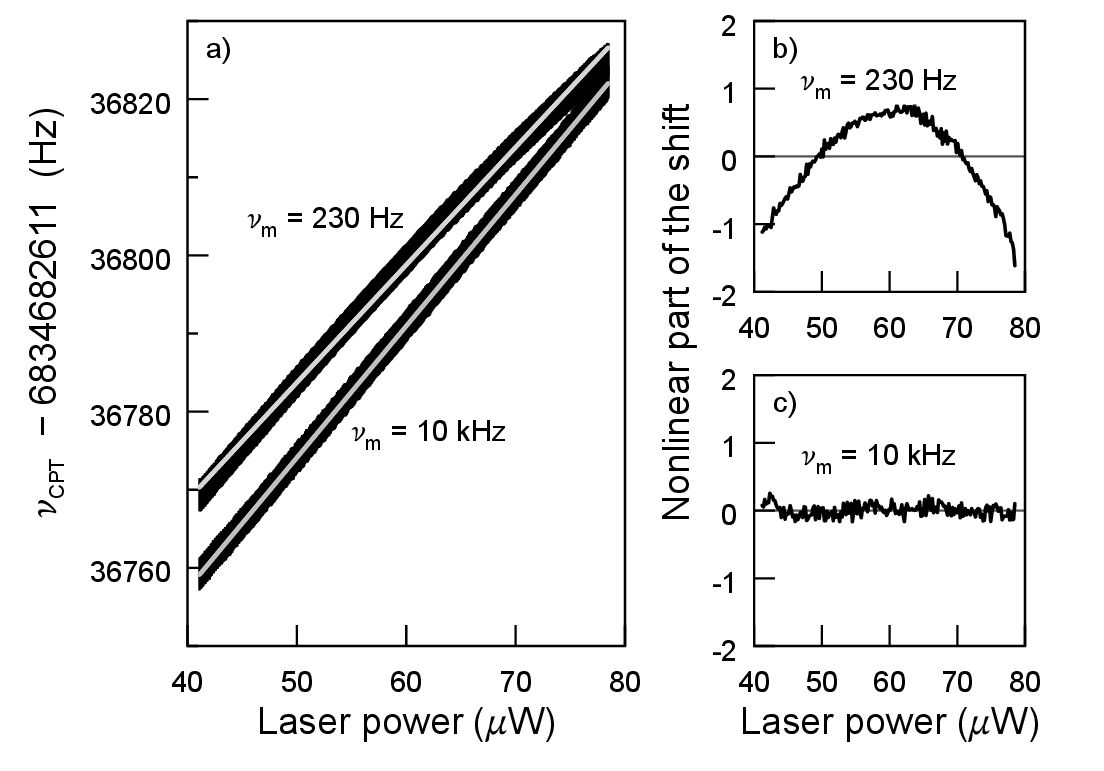}
\caption{a) Dependence of the error-signal frequency on the laser power. $\nu_m=230$~Hz corresponds to a point on the slope of the dependence demonstrated in Fig.~\ref{Pulling}. The grey lines are the linear fit. b-c) Nonlinear part of the curves presented in~a).}
\label{Nonlinearity}
\end{figure}

We obtained the dependencies of the error-signal frequency on the optical field power for two cases: when the pulling is present ($\omega_m/2\pi=230$~Hz) and when it is suppressed ($\omega_m/2\pi=10$~kHz), Fig.~\ref{Nonlinearity}a. The modulation index was $m=0.65$. The first case corresponds to a point on the slope of the dependence shown in Fig.~\ref{Pulling}. The VCXO frequency was stabilized and measured in the same way as described above, and the laser power was linearly varied using an acousto-optic modulator. As can be seen, the pulling gives a frequency shift of the order of $10$~Hz, which nonlinearly depends on the optical field intensity. Therefore, the transition to high values of $\omega_m$ provides significant reduction of $\delta_{as}$, but its nonlinear part is only near $1$~Hz; see Fig.~\ref{Nonlinearity}b. This residual shift was obtained by subtracting the linear fit from the experimental data.

\section{Discussion}

Is there a significant loss in the error-signal steepness at high modulation frequencies than when it is moderate? Our theoretical considerations demonstrate that there is a not-so-large decrease of about $1.717$ times when the best possible mixture of in-phase and quadrature signals is taken for comparison. Moreover, the use of high-frequency modulation can be more beneficial due to decreased level of low-frequency noise.

The steepness of the central dispersive curve is determined by the product $J_0(2m)J_1(2m)$, as follows from Eq.~\eqref{quadr}. So, optimization for the case of high-frequency modulation is straightforward: one should take the modulation index $m$ value $\simeq0.541$, which maximizes the product. Therefore, the high-frequency regime is convenient in the following way: the slope steepness of the central dispersive curve does not depend on the ratio of the modulation frequency to the ground state relaxation; only the value of $m$ should be tuned. Here we remind for completeness that in the case $\omega_m\rightarrow\infty$, the in-phase signal amplitude is zero in the vicinity of $\delta=0$.

In many cases, the situation of large values of modulation frequencies is called the Pound-Drever-Hall regime. However, in our case, the width of the CPT resonance also grows with $\omega_m$, its peaks become more spaced from each other. As a result, it is impossible to make the modulation frequency significantly greater than the resonance width. Therefore, it is more correct to talk about the resolution of a multi-peak structure with an increase in the modulation frequency than about the transition to the Pound-Drevere-Hall regime.

Recently we have demonstrated that the powers of the resonant components can be equalized by modulation of the injection current at single and doubled frequencies, i.e., when $j(t)=j_0+j_1\cos{\Omega t}+j_2\cos{(2\Omega t+\varphi)}$~\cite{bogatov2022control}. Here $j_0$ is the direct current, $j_1$, $j_2$, $\varphi$ are amplitudes of modulation and the relative phase shift between them. In this case the value of $\delta_{as}$ can be reduced by tuning ratio $j_2/j_1$ and the value of $\varphi$, respectively. Therefore, working at high frequency $\omega_m$ under symmetric spectrum at first glance could be seen as only reducing the slope steepness.

But we remind that there are another sources of the resonance asymmetry, for example: imbalance in populations of working sublevels due to unequal spontaneous transitions from the excited state to the ground one~\cite{sabakar2023effect}, inhomogeneity of the magnetic field, temperature gradient in the atomic cell~\cite{oreto2004buffer}, transverse inhomogeneity of the laser beam when the light shift is unsuppressed~\cite{camparo1983inhomogeneous,chuchelov2020study}, the optical field absorption~\cite{https://doi.org/10.48550/arxiv.2012.13731}. Therefore, we generalize results of work~\cite{yudin2023frequency}: an increase in frequency $\omega_m$ far beyond the ground-state relaxation rate should not only suppress the frequency pulling due to unequal powers of the resonant components, but due to all mentioned above factors leading to asymmetry of the CPT resonance.

\section{Summary}

We have considered a situation when a modulation is used to obtain the in-phase and quadrature signals in the light transmission. In this case the resonance consists of several peaks pulling the frequency of the central one under the asymmetry. The pulling drops off when peaks are resolved, i.e., when the modulation frequency $\omega_m$ significantly exceeds the relaxation rate of the ground-state coherence. For each peak of the optical field transmission there is a dispersive-shape curve in the quadrature signal, i.e., it has a multi-dispersive structure. The central curve is also affected by the frequency pulling: as we have demonstrated, its frequency dependence on $\omega_m$ vanishes when the frequency value is sufficiently large. Finally, we have demonstrated that suppression of the frequency pulling provides linear dependence of the quadrature signal frequency on the optical field intensity.

\section{Acknowledgments}
\label{Acknowledgments}

The authors receive funding from Russian Science Foundation (grant No.~19-12-00417).

\bibliographystyle{apsrev4-1}
\bibliography{references}

\end{document}